\documentstyle[draft,psfig]{mn}

  \def\etal{{\it et al.\thinspace}}
\def\ergs{\hbox{ergs}\hbox{~s}^{-1}} 
\def\kpc{\hbox{kpc}}

\begin{document}

\title[Bubbles,  Feedback  and  the   ICM]{Bubbles,  Feedback and  the
Intra-Cluster     Medium:  Three-Dimensional Hydrodynamic Simulations}
\author[Quilis et al.] {Vicent Quilis$^{1,2}$, Richard. G. Bower$^{1}$
\& Michael L.   Balogh$^{1}$\\ $^{1}$Department of Physics, University
of    Durham,       South   Road,     Durham,      DH1   3LE,     UK\\
$^{2}$email:Vicent.Quilis@durham.ac.uk\\ } \maketitle

\begin{abstract}
We use a three dimensional hydrodynamical  code to simulate the effect
of  energy  injection on  cooling flows   in the  intracluster medium.
Specifically, we compare a simulation of a 10$^{15}$ $M_\odot$ cluster
with radiative cooling only, with a second simulation in which thermal
energy is injected 31 $\kpc$  off-centre, over 64  kpc$^{3}$ at a rate
of $4.9\times 10^{44} \ergs$ for  50 Myr.  The  heat injection forms a
hot,   low density  bubble  which  quickly rises,  dragging behind  it
material  from the cluster  core.  The rising bubble  pushes with it a
shell of  gas which expands and cools.   We find the appearance of the
bubble in  X-ray temperature and luminosity  to be in good qualitative
agreement with recent  {\it  Chandra} observations of   cluster cores.
Toward the end of the simulation, at 600 Myr, the displaced gas begins
to fall back  toward the core, and  the subsequent  turbulence is very
efficient at mixing the low and high entropy  gas.  The result is that
the cooling flow  is disrupted  for  up to   $\sim 50$  Myr after  the
injection  of energy  ceases.   Thus, this  mechanism provides a  very
efficient method for regulating cooling flows, if the injection events
occur with a 1:1 duty cycle.
\end{abstract}


\section{Introduction} 
Radiative  cooling in clusters  of galaxies  makes the distribution of
gas  unstable.  As the gas in  the centre of  the cluster radiates its
internal energy,  the pressure support in  the outer  parts is reduced
and a  flow is established.  In cluster  centres, the cooling  time is
short  compared  to the  age  of  the  cluster, which   makes  this an
important  mechanism for  accreting gas onto  a  central object.   The
astrophysical  puzzle is to  understand  how the high  flow rates that
have been estimated for some clusters (up  to 2000 $M_\odot$ per year,
e.g. Allen 2000\nocite{Allen00}) can be compatible with the relatively
small  masses of  cold gas and   young stars seen  in central  cluster
objects   \cite{C+95,E+99,W+00},  or with   the  lack  of gas seen  at
temperatures      below  about  2    keV    (eg.,    Peterson  et  al.
2001\nocite{P+01}; Tamura et  al.,  2001\nocite{T+01};  Oegerle \etal\
2001\nocite{O+01}).

In smaller mass haloes, cooling flows are expected to be the mechanism
by which individual galaxies form.  However, a similar problem to that
observed in  clusters  exists, in that the  short  cooling times imply
that a large fraction of the total baryonic mass should have been able
to cool in a Hubble time.  This is in conflict with observations which
show that only a small  fraction of baryons are  now in this cold form
\cite{FHP,baryons}.

A  popular  solution to   these problems  is the  suggestion  that gas
cooling might trigger heating processes  (feedback) which regulate the
cooling flow (eg.,  White \&  Frenk 1991\nocite{WF91}).  For  example,
cooling gas is likely to lead to star  formation, which returns energy
to   the intracluster medium (ICM) via   supernovae and stellar winds;
also,   cooling  gas  accreted onto   black holes    might fuel highly
energetic, relativistic jets.  These processes might either reheat the
gas  from   the cold  phase, or  reduce   the  cooling rate, therefore
limiting   the fraction of  baryons  which   can cool;  this can  have
important consequences for the evolution of clusters and the formation
of   galaxies (eg.,  Ponman,   Cannon \& Navarro  1999\nocite{Ponman};
Valageas         \&   Silk         1999\nocite{VS99};    Bower      et
al. 2001\nocite{Bower01}).

Although there  are many physical  processes known which may be viable
sources of feedback energy, it is unknown  exactly how such injections
of energy interact with the surrounding gas.  In particular, there are
three possible outcomes  of an energy  injection event: (1) the energy
may serve to increase the total emissivity of the gas (for example, by
compressing surrounding  material),   resulting in all  of  the  input
energy radiating  away; (2) the energy  may be concentrated in a small
amount of gas that  then escapes from  the cluster core but has little
effect on the cooling  flow; (3) the injected  energy may  disrupt the
cooling  flow, either temporarily or  permanently.  Only  in the third
instance   is the mechanism  going  to  be  effective  at solving  the
problems described above; the focus of this  work is to determine what
happens to energy injected in clusters with substantial cooling flows.

Advances in computing power have recently made detailed simulations of
feedback  viable,  and  there has  consequently  been   a lot  of work
studying   the   phenomenon  with    two-dimensional    hydrodynamical
simulations.   Kritsuk \etal\  \shortcite{KPM} studied  the  effect of
supernova injecta on cooling flows on galactic scales, and showed that
overlapping  supernovae   within   the  galaxy  created   a  turbulent
convection zone that  effectively mixed  gas   in the inner   regions,
reducing the entropy profile  and the central  cooling rate.  Churazov
\etal\ \shortcite{Churazov} consider  the specific case of  the galaxy
M31,  and  study in detail the   dynamics of buoyant bubbles  within a
model galaxy, and their influence on the surrounding medium.  However,
their simulations do not include radiative  cooling, and are therefore
unable to address the ability of such bubbles to influence the cooling
rate.   Reynolds, Heinz \&   Begelman \shortcite{RHB} investigated the
behaviour of   supersonic  energy sources  in cooling   flow clusters,
concentrating on the interaction of radio  jets with the intra-cluster
medium.

All   of the recent    simulations discussed above   find that  energy
injection results in buoyant   bubbles which displace and  disturb the
surrounding  gas.   In this  paper,  our purpose is  to investigate in
detail the  effect  that energy  injection into  the  ICM has  on  the
cooling  rate  of the surrounding  gas.    We present  the first fully
three-dimensional  simulations  which  include  radiative  cooling and
sporadic energy injection  on  cluster scales.  The  three dimensional
nature   of the simulations   is   essential to ensure that  realistic
convective    flows are  established,   and  turbulent   mixing of the
multiphase gas is  properly modelled.  In \S\ref{sec-sims}  we present
the details  of  the numerical model,  and our  method for  simulating
energy  injection.  The results of  two simulations, one  with and one
without  energy injection,  are presented   in \S\ref{sec-results}.  A
quantitative analysis of the  results, and qualitative comparison with
{\it Chandra} observations  are presented in \S\ref{sec-discuss}.  Our
conclusions are summarized in \S\ref{sec-concs}.

\section{Simulations}\label{sec-sims}

\subsection{Numerical Code}
We use the  three dimensional Eulerian fixed-grid  hydrodynamical code
described  in Quilis \etal \shortcite{Quilis}.    The code uses modern
{\it  high-resolution shock-capturing}  (HRSC) techniques,  which  are
specially designed to integrate hyperbolic systems of equations as the
hydrodynamic equations.  The HRSC techniques have important advantages
over some other techniques.   The practical implementation of the code
has   four  key ingredients:  i)   conservative  formulation; that is,
numerical  quantities are conserved up to   the numerical order of the
method, ii) the reconstruction procedure, which  allows to recover the
distribution  of the  quantities inside the  computational cells, iii)
the Riemann solver, which  solves the evolution of  discontinuities at
cell interfaces, and iv) the time advancement, which is designed to be
consistent with  the conservation  properties.  The main  advantage of
HRSC codes  is the use  of  Riemann solvers to  compute the  numerical
viscosity  needed to   solve  the  hydro  equations.  This   numerical
viscosity is given  internally by the  method and does not require any
guess at  the  form of  the  viscosity  (the  so called   ``artificial
viscosity'' required by other methods).  The numerical viscosity given
by Riemann  solver based methods is small  and therefore these methods
can resolve strong  shocks  extremely well  (typically in one   or two
cells) as the diffusion is  reduced dramatically compared with methods
based on  other prescriptions   for  the numerical  viscosity.   Other
important advantages of HRSC  techniques are that  they work very well
in low density  regions, and they are of  high-order in smooth regions
of the flow.

The nature of the   feedback process presents important  challenges to
the numerical code.   Specifically,  the sporadic injection   of large
amounts of energy  in  a very small  volume  could produce shocks  and
large jumps in the hydrodynamical quantities  which must be accurately
resolved.  Large  jumps  in  density  are also   produced by turbulent
mixing and radiative cooling, as well as the propagation of the bubble
which generates  sound waves.  Our   hydrodynamical code was specially
designed for an accurate treatment of fluid dynamical processes and is
extremely  good  in  dealing   with shocks,   strong  discontinuities,
turbulent regions, and  low density regimes.  Therefore,  it is a good
tool to tackle the problem addressed in this paper.

The system of  hydrodynamic  equations needs a   equation of state  in
order to be closed.  We adopt the equation of  state for an ideal gas,
$p=(\gamma-1)\rho\epsilon$, where $p$   is the pressure,  $\gamma$ the
adiabatic exponent, $\rho$  the density,  and $\epsilon$ the  specific
internal energy.  Considering the range  of temperatures and densities
of the ICM, we model the ICM gas as a non-relativistic monatomic fluid
with  $\gamma=5/3$.   Although  this   assumption  is  reasonable   --
especially at  the  resolution of our simulations  --  it neglects the
role  of magnetic fields  which might lead  to more long-lived bubbles
(Fabian et  al.  2001).    Magnetohydrodynamic interactions  cannot be
addressed with our present code.

\subsection{Initial Conditions}
We run two simulations of a $10^{15}M_\odot$ cluster,  using a grid of
$256^3$ cells in a  cube 1 Mpc per side,  corresponding to a cell size
of $3.9  \kpc$.  The  simulations are carried   out on an Origin  2000
parallel computer, and follow the  cluster evolution over a  simulated
time  of  900 Myr.  We  assume  a Hubble constant   of $H_\circ=70$ km
s$^{-1}$.

The initial cluster profile   is constructed assuming  that isothermal
gas  sits  in  a $10^{15}M_\odot$   dark    matter potential well   in
pressure-supported  hydrostatic  equilibrium.   For  the  dark  matter
potential we assume the modified-NFW profile  found in high resolution
simulations \cite{NFW-Moore,Lewis-sim}.  The  virial radius, 2.55 Mpc,
is    computed  using   the   formalism   described  in  Babul  \etal\
\shortcite{Babul2}, assuming     a  cosmology with $\Omega_\circ=0.3$,
$\Lambda=0.7$,   and    $H_\circ=70$   km  s$^{-1}$.     We  assume  a
concentration parameter $c=4$, which implies that  the ``core'' of the
potential is  25\% of   the virial  radius,  or 640$\kpc$.     The gas
temperature  is set to the virial  temperature  of the halo, $kT=4.75$
keV.  We assume that the mass fraction of gas within the virial radius
is given by  $\Omega_b/\Omega_\circ$,  where $\Omega_b=0.039$  is  the
universal baryon density, as  determined from deuterium  abundances by
Burles, Nollett  \&   Turner \shortcite{BBN2}.  The  bolometric  X-ray
luminosity of this model cluster is $4.9\times 10^{44}$ ergs s$^{-1}$,
where the emissivity is  computed using the model  of Raymond, Cox  \&
Smith   \shortcite{RCS}.  Model clusters   based on  these assumptions
agree quite well  with   observed scaling relations of   dynamical and
X-ray properties of massive clusters, though they are less suitable in
low mass systems   \cite{Babul2}.   In order  to  break  the spherical
symmetry of this model, we introduce randomly 10\% fluctuations in the
gas density.

The   boundary  conditions    are   treated  with    an inflow/outflow
condition.  Practically,  this means that  the  values at the boundary
cells are determined by copying the values  of their inner neighbours.
Our  results  are insensitive  to   the boundary conditions,  however,
because the energy  injection event occurs far  from any boundary, and
the simulation is  not run long  enough for effects at  the edge to be
felt in the centre.  We have verified that other reasonable choices of
boundary conditions   do   not have    significant  effects  on    our
conclusions.

\subsection{Energy injection}\label{sec-inject}
We   choose to inject a   quantity  of energy in   a  region which  is
displaced from the cluster centre; this choice  is conservative in the
sense that it  is  less likely   to  be successful at  disrupting  the
cooling flow than  if we had deliberately  targeted the heating on the
lowest  entropy gas.  This   will  ensure that we  distinguish between
scenarios (2) and (3) discussed in  the introduction: will a bubble of
energetic gas not  centred on the cooling flow  have  any influence on
that flow?

Our model is  not intended to investigate the  interaction of  the AGN
jet  with the surrounding material \cite{RHB},  so we simply model the
heating process by  injecting  energy centered   on  a cell  which  is
displaced 8 cells ($31 \kpc$) from the  cluster center. The heating is
distributed over the  surrounding pixels using a spherical  (SPH-like)
kernel with smoothing length 8 cells.  The heating  rate was chosen to
be of the same  order as the cooling  rate of the cluster:  $4.9\times
10^{44} \ergs$; this is expected to be typical of the energy injection
from  radio jets   (e.g.    Owen, Eilek \&  Kassim  2000\nocite{OEK}).
Heating at a much larger rate results  in an explosive detonation with
much of the energy  being radiated in a dense  shell (Reynolds et  al.
2001); at  a lower  rate the  heat   input is unlikely   to affect the
cluster cooling flow.

The  heat input is continued for  50 Myr, by which   time a near empty
bubble has been formed in the ICM. It is then switched off so that the
energy can be distributed around the cluster. This is motivated by the
desire to model a scenario in which the AGN (or other feedback source)
is ``woken-up''  by gas deposited  in the cooling flow.  This disrupts
the flow which turns off the supply of fresh material and hence limits
its lifetime.

\subsection{Integrated quantities from two dimensional slices}\label{2Dv3D}

The simulations shown in this paper are fully 3D. Although we seed the
initial conditions with random  fluctuations and the source  of energy
is located off-centre, the results of our simulation show a high axial
symmetry in their gross details.

This degree  of symmetry has beneficial consequences  on the amount of
computational   resources  needed,    as we    can  compute integrated
quantities  such as entropy,   mass   or luminosity from 2D    slices,
avoiding  the need to store huge  3D  outputs.  Consequently, a better
tracking of the time evolution of these quantities is also possible as
a larger number of outputs can be saved.

\begin{figure}
\begin{centering}
\psfig{file=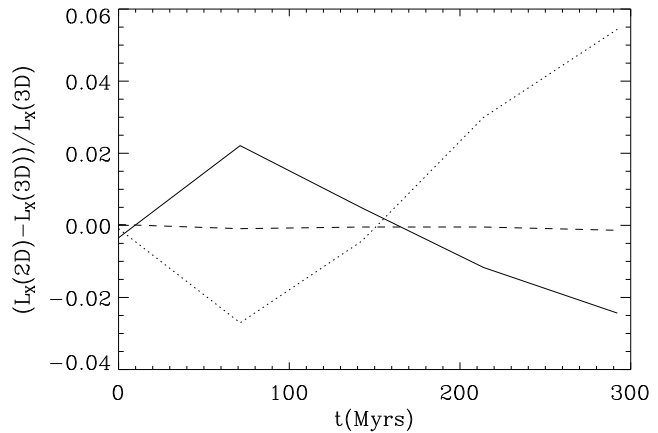,width=9cm}
\end{centering}
\caption{The relative  error in  the  X-ray luminosity,  $L_X$ -- when
computed  from two   dimensional   slices  assuming   the  axisymmetry
approximation   -- as a function  of  time.  The  three lines show the
results for three orthogonal two dimensional slices extracted from the
full three dimensional data set. The relative errors are computed with
respect to the $L_X$ luminosity computed using the  whole 3D data box,
for five different outputs.}
\label{fig-lx-error}
\end{figure}

The axisymmetric  approximation necessarily introduces some error when
computing integrated quantities from the 2D slices instead of from the
full  3D  data set.  In order  to  quantify  this, and  to justify the
validity of the approximation, we have  computed the relative error of
the  X-ray   luminosity computed  from three   orthogonal 2D slices --
extracted from   the  3D full data --    assuming axial symmetry, with
respect  to  the  X-ray luminosity  computed  from the  whole 3D  box.
Figure \ref{fig-lx-error} shows the evolution of the relative error as
a function of time for 5 outputs where we  have the full 3D output. In
any  case errors are larger   than  6\%.   Other quantities, such   as
density, show  even smaller  errors.  The  axisymmetry of  the problem
does not obviate the use of 3D simulations  as mixing processes at the
gas interface,  for example, may  be  more correctly modelled,  though
still symmetric.

\begin{figure*}
\begin{centering}
\psfig{file=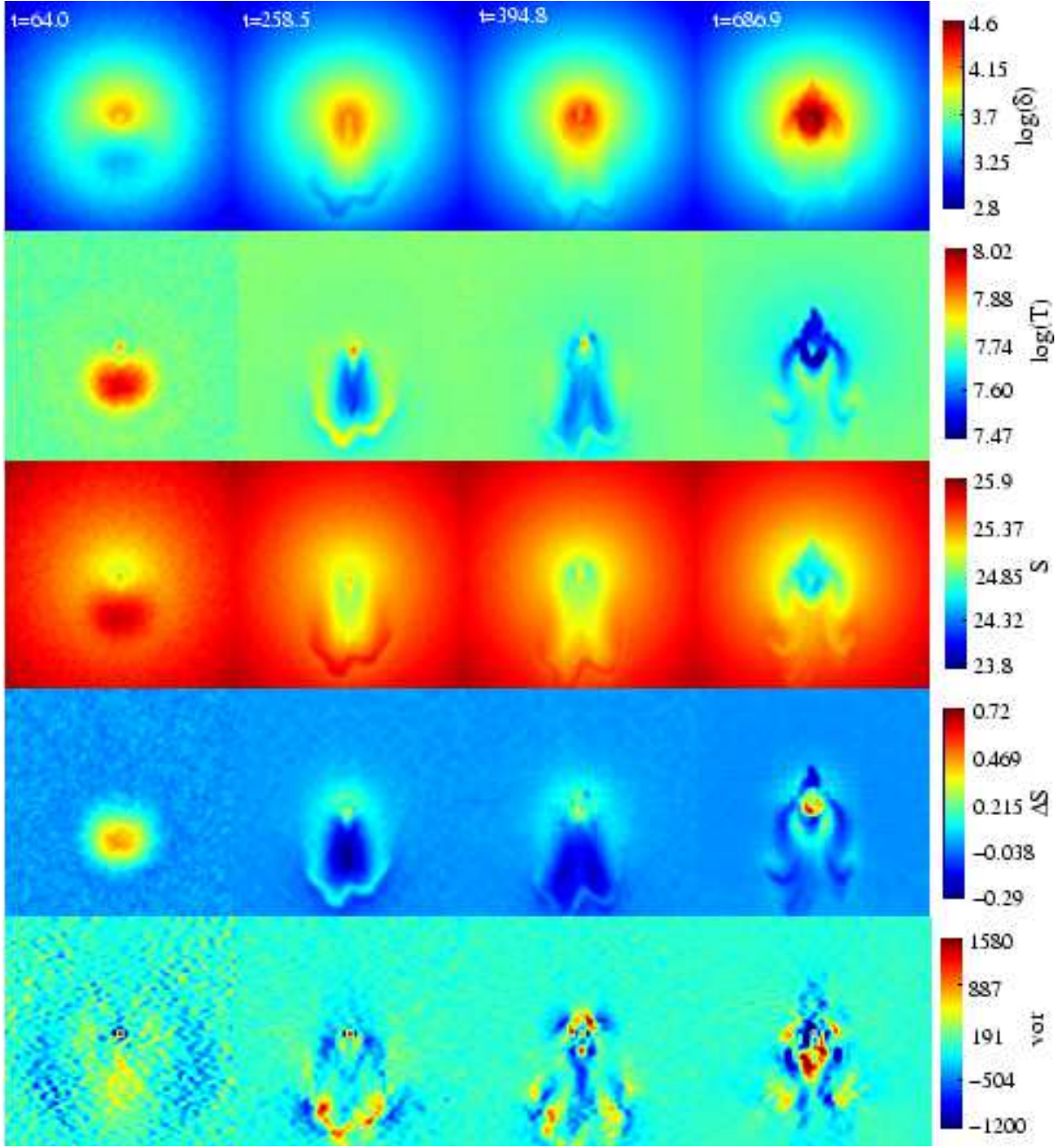,width=16cm}
\end{centering}
\caption{A time  sequence of two  dimensional slices taken through the
fully 3 dimensional simulations with heat injection.  The box shown is
$250 \kpc$ on each side; the full simulation  is 1 Mpc.  The different
rows show evolution in the  ICM (a)~density contrast; (b)~temperature;
(c)~entropy;  (d)~differential   (compared with     the   cooling-only
simulation) entropy; (e) vorticity.   The panels are labeled  with the
time in Myr since the start of the simulation. \label{fig-slices} }
\end{figure*}

\section{Results}\label{sec-results}

\subsection{Cooling Simulation}

The first   simulation  that we run   includes  radiative cooling, but
without any extra energy input.   The flow quickly becomes established
(at 50 Myr) and  continues at a steady   rate for the duration  of the
simulation.   The cluster collapses  slowly as  the energy is radiated
(see Figure \ref{fig-evoln}); as matter flows into the central region,
its luminosity increases,     and  the mass-weighted     mean  entropy
falls. These effects, including the steepening  density profile of the
cluster core,  have been   seen  previously (eg.,  Knight  \&  Ponman,
1997\nocite{KP97}; Lewis  et   al.  2000\nocite{Lewis-sim}; Pearce  et
al. 2000\nocite{Pearce}; Kritsuk et al., 2001\nocite{KPM}).

\subsection{Feedback Simulation}

In this simulation, energy is injected into  a single cell over 50 Myr
as  described in \S\ref{sec-inject}.  We display  the evolution of the
gas properties in a two-dimensional  slice in Figure \ref{fig-slices}.
The  top three rows show   the gas density  contrast, temperature  and
entropy, at 4 different times.   The full three-dimensional simulation
is approximately   axisymmetric  about     the  $y$-axis   of    these
slices.\footnote{A movie displaying the formation and evolution of the
bubble       can        be    seen       at      the        web   site
http://star-www.dur.ac.uk/$\sim$quilis/movies/bubble\_den.mpg.     The
movie shows  the  time evolution  (in units of   $10^6$ years)  of the
logarithm of the overdensity.}

The  energy injected  heats  a    relatively  small amount   of  mass,
$3.5\times 10^{10} M_\odot$,  to $10^8$ K, creating  a  hot but almost
empty  bubble.  In effect, the bubble  is generated by a subsonic wave
expanding  into the cooling  ICM (a sonic boom   in the terminology of
Reynolds  et al.   2001\nocite{RHB}).  In  the simulation, the  energy
injection is sufficiently  weak that the  bubble expands subsonically,
pushing  gas out  of the cluster   centre.   Since this gas  is moving
adiabatically from a high pressure region to one of lower pressure, it
expands and cools.  This results in  a shell of  gas around the bubble
which is {\it cooler} than the  surrounding material.  As we elaborate
upon in    \S\ref{sec-Xray},  this is qualitatively  similar   to {\it
Chandra}   observations     in   some  clusters   (e.g.     Fabian  et
al. 2000\nocite{Fabian-Perseus},  2001\nocite{Fabian-A1795}).   Unlike
the simulations of Reynolds  \etal\ \shortcite{RHB}, the cold shell is
not due to the displacement  of already cold  gas, although the result
is similar.

Once the heat source is turned off (at 50  Myrs), the bubble ceases to
expand, but does not collapse. The buoyancy  of the bubble compared to
the surrounding material takes over as the  dominant force, pushing it
outward from the centre of the  simulation.  This causes the bubble to
adopt a cap-like geometry; roughly a semi-spherical  shell of gas with
a low density, low pressure interior.  This is similar to the geometry
found in the  simulations of Churazov \etal\  \shortcite{Churazov}; in
their case,  the  force of the  rising  material behind  the bubble is
strong enough to punch through and cause the shell  to become a torus.
The  surface of the  shell   in our simulation experiences   turbulent
mixing  with  the  surroundings.   The shell   rises until its entropy
becomes   comparable  to the surrounding   medium,   at which point it
dissolves into the surrounding  ICM. This occurs  well within the core
region of the cluster ($<250 \kpc$).

In order to better illustrate the effect of  the bubble on the entropy
distribution of the cluster, we have taken a slice through the cluster
showing the  difference    between the   entropy   of the   `feedback'
simulations  and the `cooling'   simulations.  This  is  shown in  the
fourth row (d) of Figure \ref{fig-slices}; these plots show the effect
of the bubble more clearly by subtracting away the entropy gradient in
the  initial profile and the   entropy decrease due   to cooling.  The
initial, localised   entropy   increase resulting    from  the  energy
injection is  clearly  seen in  the   first panel.   Subsequently, the
entropy difference of the  bubble becomes less  apparent: as it rises,
the surrounding gas has higher entropy and the  contrast of the bubble
declines. Eventually, the heated material reaches a  point at which it
mixes with the surrounding material and disappears.

As a result of the low pressure region created behind the bubble as it
rises, a plume of  low entropy material (darker  blue colour) is drawn
out of the  core  (see  also Churazov \etal\   2001\nocite{Churazov}).
This gas, although not heated  directly, becomes mixed into the higher
entropy gas surrounding the  cluster, slightly depressing the  entropy
on  scales of 50--100$\kpc$.  But the  entropy of the very central gas
(within $10  \kpc$) is higher in  the feedback simulation, relative to
the  cooling-only simulation.   This  is because,  in the cooling-only
simulation, the dense core gas has radiated  its energy and dropped in
entropy;  this  cooling has been   prevented  in the  simulation  with
feedback, resulting in a relatively higher core entropy.

The core is significantly  disturbed by the  rising of the bubble, and
turbulent mixing  of  gas  in  this region   becomes  efficient.    To
demonstrate the efficiency of mixing, we show the vorticity of the gas
in  the bottom  panel  (e)  of Figure   \ref{fig-slices}.  During  the
initial formation of the bubble, the vorticity is low, and largely due
to the density fluctuations in  the initial conditions.  As the bubble
rises, mixing occurs  predominantly at the edge  of the shell.  By the
end of the  simulation, some of  the gas drawn  out by the  bubble has
begun to fall  back onto the  cluster centre, dragging with  it higher
entropy   material  from the outer    regions  into the  now-turbulent
core. The cluster centre is then a region of high vorticity, and plays
an   important role in    mixing the high and  low   entropy gas in an
irreversible process, thus generating a net increase in the entropy of
the cluster core. It is also at this  time that cooling in the central
regions,  which had  been  disturbed by the  initial  expansion of the
bubble, once again becomes strongly established.

\section{Discussion}\label{sec-discuss}
\subsection{Numerical Analysis}

\begin{figure}
\begin{centering}
\psfig{file=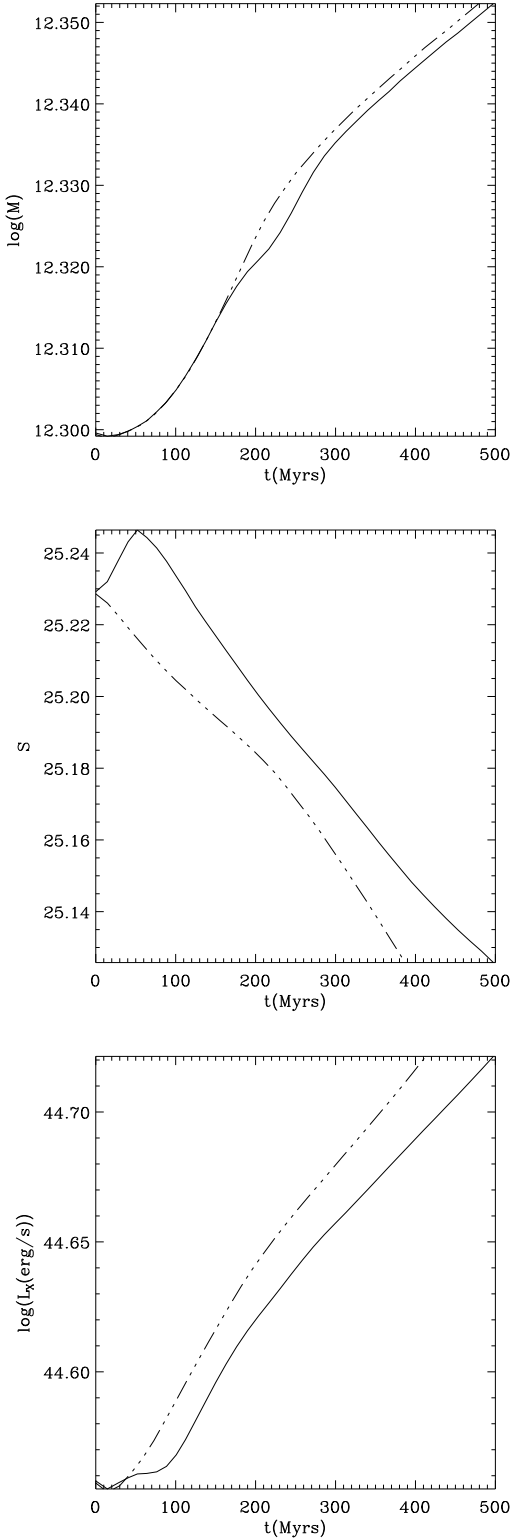,width=7cm}
\end{centering}
\caption{Evolution of average quantities  within a radius of $200\kpc$
in  the simulations:    (a)~enclosed mass; (b)~mass  weighted  average
entropy; (c)~integrated   X-ray luminosity.   Dashed  lines show   the
`cooling  only'  simulation, while  solid  lines  show the  `feedback'
simulation.
\label{fig-evoln}
}
\end{figure}

In this section we look at the  evolution in the integrated properties
of  the model cluster.  We focus  on a sphere  centered on the cluster
that encloses all of the bubble  material at 500 Myr (corresponding to
a radius of $200$ kpc).  Figure \ref{fig-evoln} shows the evolution of
this enclosed  mass, the mass-weighted entropy  and the  luminosity of
the  system.  These quantities are computed  from  the two dimensional
slices shown  in Figure \ref{fig-slices},   using the approximation of
axisymmetry,  because  of  the limited  number   of large, fully-three
dimensional outputs (see \S\ref{2Dv3D}).

The  mass enclosed within  the boundary evolves  similarly  in the two
simulations, increasing by about 10\% due to  the infall of gas as the
densest core gas radiates its energy.  At late times in the simulation
with heating, a small amount of mass , $1.4\times 10^{10} M_\odot$, is
pushed out of the region by the rising plume of gas.

Concentrating  on the mass-weighted mean  entropy  of the system,  the
initial effect of the heat injection can be  clearly seen.  This small
rise in entropy is not sufficient to prevent the  cooling of the core,
which subsequently continues at a rate similar to  that in the cooling
only simulation.   The primary effect  of  the bubble is  therefore to
delay the entropy evolution   of the the  system.   At the end of  the
heated  simulation, the  average entropy is   the same as that  of the
unheated model 100 Myrs earlier.

We set out to distinguish between three possible fates of the injected
energy:   1)  the energy  is quickly  radiated  away, resulting  in an
increase in X-ray  luminosity; 2) the  bubble escapes from the cluster
core  without affecting the  cluster core,  in  which case the rate of
energy radiation is  unchanged; or 3)  the heating event disrupts  the
cooling flow, reducing the  X-ray luminosity.  The dominant effect  in
our  simulations  is  scenario  3):  the  injected  energy temporarily
disrupts  the cooling flow, but  is not immediately radiated away, nor
does  it escape the  cluster core.  In  the  cooling-only run, a total
$9.2\times10^{60}$ ergs have   been    radiated by  the  end  of   the
simulation, compared with only $8.8\times10^{60}$ ergs in the run with
energy injection.  The difference is  $3.46 \times10^{59}$ ergs.  This
is 55\% of the total energy injected. The total  energy of the cluster
 differs between the  two simulations by 1.55 times the  total energy
injected.
   
Since the cooling  flow  is  re-established  about 50  Myr after   the
heating source is  shut off, this mechanism  can only be  effective at
substantially reducing the  cooling rate over a  Hubble  time if there
are multiple heating events.  Observations show that about 40 per cent
of central  cluster galaxies show strong  line emission, indicative of
either star formation   or nuclear activity, which could   potentially
heat the surrounding gas \cite{JFN,HBBM,MOC,C+99}.  This suggests that
a  duty cycle of 1:1 for  the heating events  is not unreasonable.  In
our cluster model, if such 50 Myr heating periods were initiated every
100  Myr, a total energy  of 6.28$\times10^{70}$ keV would be injected
into the cluster within  a Hubble time  (13 Gyr).  This corresponds to
0.9  keV per  baryonic particle  within the  virialized  radius, which
compares well with the energy  required to explain   the slope of  the
observed         X-ray   luminosity-temperature            correlation
\cite{Wu,Ponman,entropy}.

\subsection{X-ray Observations}\label{sec-Xray}

\begin{figure*}
\begin{centering}
\psfig{file=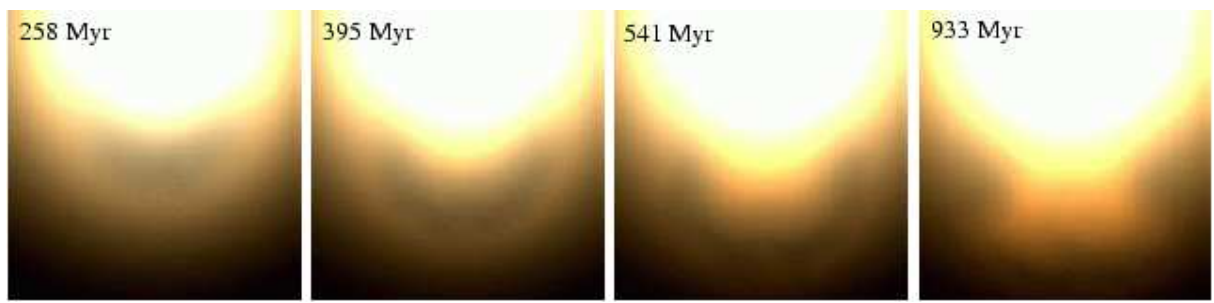,width=16cm}
\end{centering}
\caption{Mock X-ray images of a cluster generated from our simulation.
A sequence of  output times from  258~Myr  to 933~Myr are shown.   The
intensity of the image  shows  the X-ray   surface brightness of   the
material in the cluster core, while  the colour of the image indicates
the luminosity weighted temperature of  the emission.  The temperature
scale  according to the  color palette is red  (0.5-2 KeV), green (2-6
Kev), and blue  (6-10 keV).  The panels  are labelled with  the output
time.  The intensity has been  scaled to  emphasise the appearance and
visibility of  the  bubble; the same scaling  is  used for all  of the
panels.  The images show an region that is $120 \kpc$ on a side.}
\label{fig-Xray}
\end{figure*}

Recent {\it Chandra} observations have provided empirical evidence for
bubbles   in  the  ICM   \cite{Fabian-Perseus,McNamara}.    In  Figure
\ref{fig-Xray}  we show the  projected  X-ray emissivity of the heated
cluster simulation  as  a time  sequence from   258 to 933  Myr.   The
intensity  of the  image reflects the  surface   brightness across the
cluster as it would  appear in soft  energy X-rays, and is created  by
projecting the X-ray emissivity  through the volume of the simulation.
The colour the illustrates the luminosity weighted temperature in each
region.

At 258~Myr, a  warm, low emissivity bubble is  just visible.   The low
contrast     of the bubble  reflects  the    large contribution to the
emissivity made by the gas in front of and  behind the bubble. At this
time, our   simulation already has a   steeply  rising density profile
towards  the centre, so the  central  regions  are  saturated in  this
image.

As the bubble  becomes thinner and more  shell-like, it becomes better
defined  in the  X-ray image (eg.,   the  central panels show  395 and
541~Myr). The material surrounding the  bubble has similar temperature
to the undisturbed  cluster gas, so that  shell surrounding the bubble
cannot  be seen directly.  This is  in good qualitative agreement with
the observations of \cite{Fabian-Perseus}.   Because  of the high  gas
density in the  core of our simulation,  the plume of material that is
drawn  out  of the core   is easily  visible as   a distortion of  the
symmetry of the X-ray distribution.
 
The final panel shows the  cluster at a late  time (933 Myr).  At this
time,  the bubble  is no  longer  visible, having dissipated into  the
surrounding ICM. The  disturbance of the  cluster remains visible as a
plume  of cooler  gas.  Despite  the  sharp contrast  in density, this
material   is in pressure   equilibrium with  its surroundings.  Sharp
features  such  as  this are  reminiscent  of  the  surface brightness
discontinuities seen in the cores of some clusters, for example A2142

\section{Conclusions}\label{sec-concs}
We  have  presented results  from  the  first fully three  dimensional
simulations of  feedback in a  cluster model, including the effects of
radiative   cooling.   The   heating model   has   been  successful at
regulating the  cooling of the  ICM.  We  have  seen  that the  energy
injected  is not immediately radiated  and that, although only a small
amount of gas is  involved in the  formation of the bubble, its effect
is felt throughout the cluster core.  The aim of these simulations has
been to  quantify the effectiveness of this  form of energy injection,
and we summarize our findings as follows:

(1) The effect of  the delay to the cooling  lasts $\sim 50$ Myr after
the initial energy  injection event finishes,  i.e., an amount of time
equivalent to  the time over  which energy was supplied.  Therefore, a
1:1 duty cycle of heating events would be very effective in regulating
the radiative cooling process.

(2) There is  still a net  radiation  of energy in  both  simulations.
Measuring energy  relative  to the cooling-only simulation  shows that
the cooling flow has been disrupted, and the  injected energy has been
efficiently mixed into the  surrounding ICM: the radiated energy never
exceeds the energy  radiated in the ``no feedback''  case, and none of
the heated bubble material is convected out  of the cluster.  Although
our   simulations  have not  directly   reduced the   flow  rate  over
timescales of $\sim 500$ Myr, the efficiency  and timescale over which
the  cooling   flow is disrupted  makes  it  seem likely that multiple
events will be very effective at regulating the cooling rate.

Three  dimensional simulations  are   an  essential component of   the
present model,   despite   the  axisymmetry of   these   results.   In
particular, the  modelling of  turbulence,  which  is crucial  to  the
mixing of high and low entropy gas, could not be properly described by
a two   dimensional code.  A   more detailed description  of  the host
cluster may   also have important  consequences  on  this  model.  For
example,   modelling  the  expected   mass   substructure, distributed
asymetrically  in small orbiting  lumps, may increase the efficency of
energy    mixing   by  helping  to    destroy   the  pressure-confined
bubble. Accurate modelling of the these  processes will requires three
dimensional simulations such as the one we have presented here.

The next step in this work is  to simulate multiple energy events over
cosmological timescales, with a  full hierarchical merger-history.  It
will also be interesting to see if this mechanism is equally effective
on galaxy scales, and thus able to limit the fraction of baryons which
cool  globally  to  a number which  is  in  agreement  with  the tight
observational constraints \cite{baryons}.

{\bf Acknowledgments}.   We  would like to  thank  Sebastian Heinz for
useful  discussions.      VQ   is  a   Marie   Curie     Fellow (grant
HPMF-CT-2000-00052).   MLB  acknowledges support from a  PPARC rolling
grant for extragalactic cosmology at Durham.  Simulations were carried
out as part of the Virgo consortium on COSMOS an Origin 2000.

\bibliographystyle{astron_mlb} \bibliography{ms}

\end{document}